\documentclass[aps,prd,showpacs,twocolumn,superscriptaddress,%
floatfix,preprintnumbers]{revtex4}

\usepackage{epsfig}
\usepackage{amssymb}

\newcommand{\eg}{\textit{e.g.}}
\newcommand{\ie}{\textit{i.e.}}
\newcommand{\etal}{\textit{et al.}}
\newcommand{\ppbar}{$p\bar{p}$}
\newcommand{\Lepto}{\textsc{lepto}}
\newcommand{\Pythia}{\textsc{pythia}}

\begin{document}


\title{Diffractive Higgs bosons and prompt photons at hadron colliders}

\preprint{TSL/ISV-2002-0266}

\author{R.~Enberg}
\affiliation{High Energy Physics, Uppsala University, 
Box 535, S-75121 Uppsala, Sweden}

\author{G.~Ingelman}
\affiliation{High Energy Physics, Uppsala University, 
Box 535, S-75121 Uppsala, Sweden}
\affiliation{Deutsches Elektronen-Synchrotron DESY, 
Notkestrasse 85, D-22603 Hamburg, Germany}

\author{N.~T\^\i mneanu}
\affiliation{High Energy Physics, Uppsala University, 
Box 535, S-75121 Uppsala, Sweden}

\begin{abstract}
Models for {\em soft color interactions} have been successful in describing and
predicting diffractive hard scattering processes in $ep$ collisions at DESY
HERA and \ppbar \ at the Fermilab Tevatron. Here we present new comparisons of
the model to recent diffractive dijet data, also showing good agreement. The
topical issue of diffractive Higgs production at the Tevatron and LHC hadron
colliders is further investigated. For $H\to\gamma\gamma$ the irreducible
background of prompt photon pairs from $q\bar{q}\to\gamma\gamma$ and
$gg\to\gamma\gamma$ is always dominating, implying that higher branching ratio
decay modes of the Higgs have to be used. However, such prompt photons can be
used to test the basic prediction for Higgs production since
$gg\to\gamma\gamma$ involves a quark loop diagram similar to $gg\to H$. 
\end{abstract}

\pacs{14.80.Bn, 12.38.Lg, 13.85.Rm}
\maketitle


\section{Introduction}

There has recently been much interest
\cite{diffrHiggs,PRL-Higgs,KMR,Cox,DeRoeck:2002pr,Khoze:2002py} in whether the
Higgs boson could be easier observed, or even discovered, in diffractive
scattering events at hadron colliders, {\it viz.}\ \ppbar \ at the Fermilab
Tevatron or $pp$ at CERN's future Large Hadron Collider (LHC). The underlying
idea is that the lower hadronic activity in such events with large rapidity
gaps should improve the possibility to reconstruct the Higgs from its decay
products. One crucial issue is whether the cross section for diffractive Higgs
production is large enough to give an observable rate. Theoretical predictions
based on different models vary by orders of magnitude and it is therefore
important to check and constrain these models in various ways. 

\begin{figure}[htbp]
\begin{center}
\begin{tabular}{cc}
\epsfig{height= 15mm,file=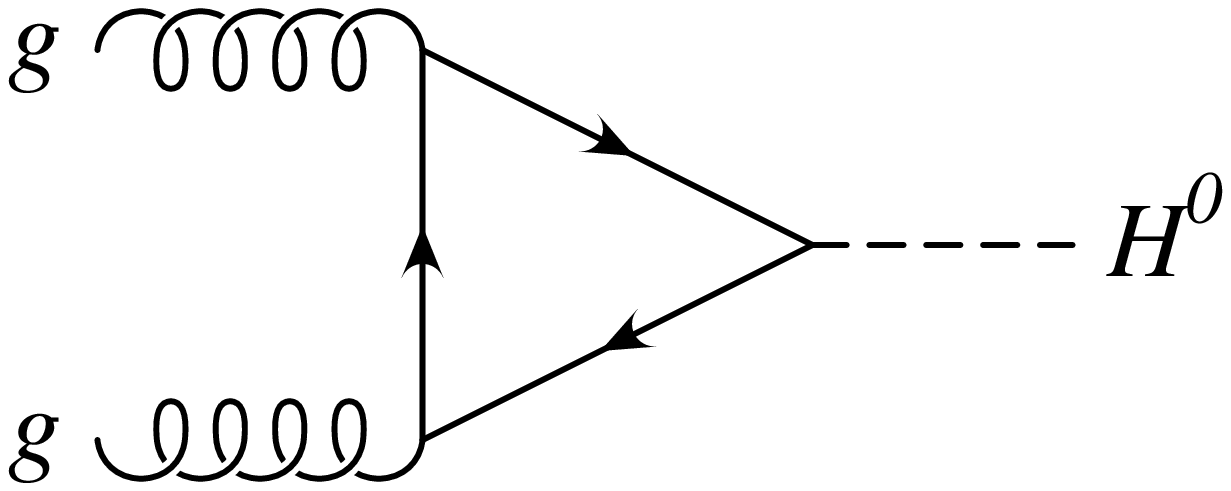,clip=}
&
\epsfig{height= 15mm,file=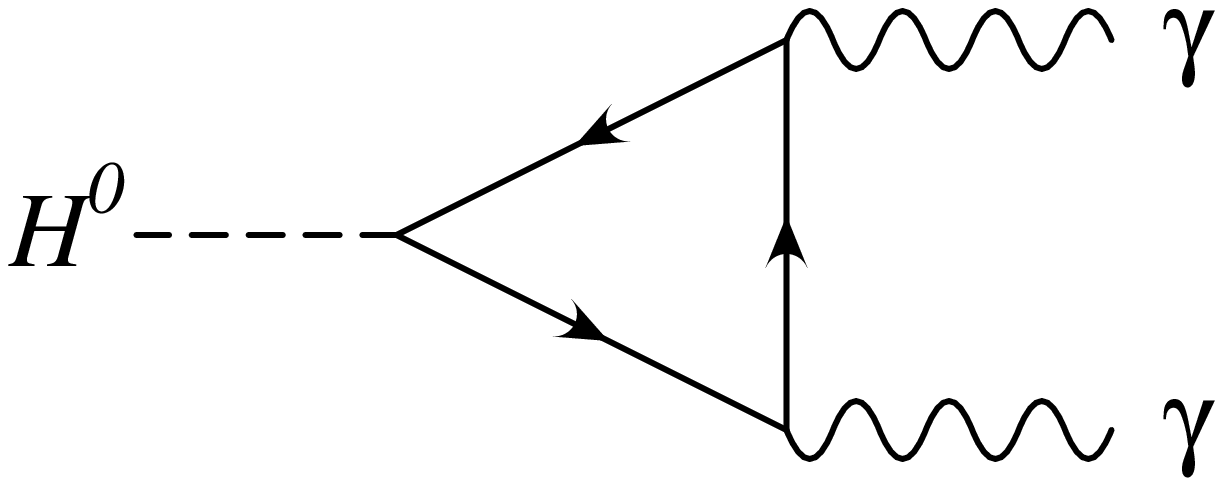,clip=}
\\
(a) & (b) \\[3ex]
\epsfig{height= 15mm,file=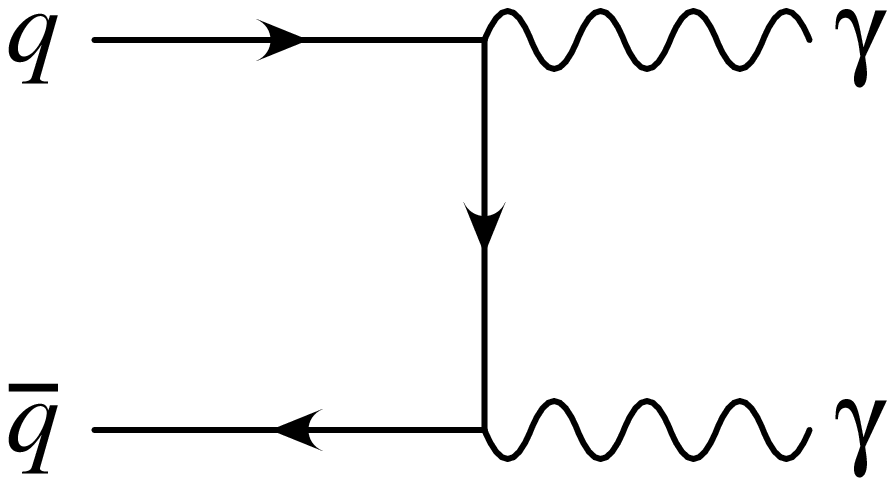,clip=}
&
\epsfig{height= 15mm,file=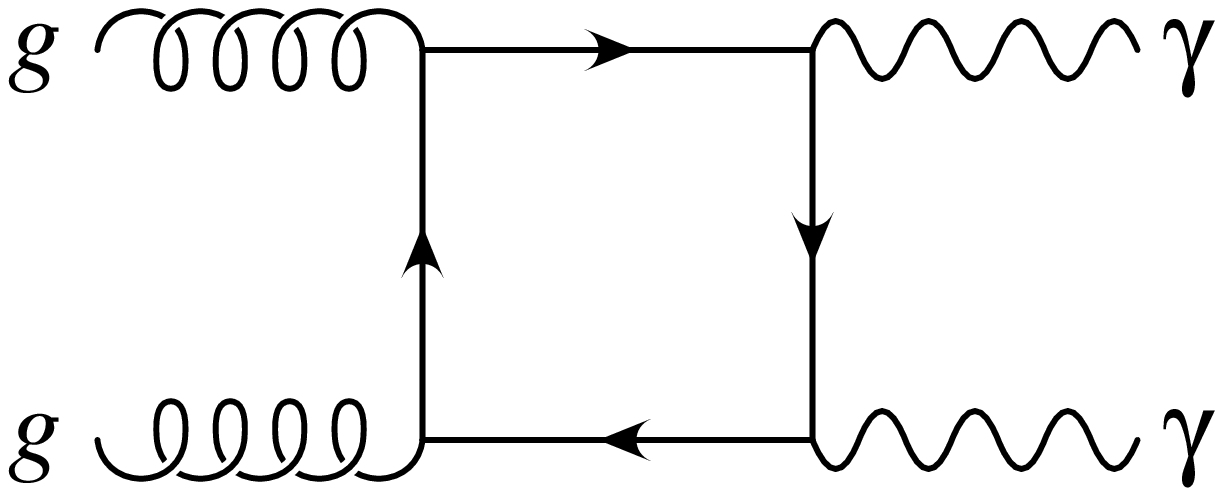,clip=}
\\
(c) & (d) 
\end{tabular}
\caption{Higgs production in (a) gluon-gluon fusion via a quark loop and 
decay (b) via a quark or $W$ loop into a photon pair. 
Prompt photon pair production in (c) leading order quark-antiquark annihilation
and in (d) higher order gluon-gluon fusion via a quark loop.}
\label{fig-diagrams}
\end{center}
\end{figure}

Another crucial issue is which decay mode of the Higgs that is considered.
Although $H\to\gamma\gamma$ (Fig.~\ref{fig-diagrams}b) has a small branching
ratio, its very clear experimental signature with two high-$p_\perp$ photons
makes it interesting and often considered for a Higgs of intermediate mass. 

In this note we will address both these crucial issues. Our prediction \cite{PRL-Higgs} for the diffractive Higgs cross section is given additional credibility by further experimental checks supporting our soft color interaction (SCI) model. The $H\to\gamma\gamma$ channel is further considered by investigating the background from prompt $\gamma\gamma$ production. The process
$gg\to\gamma\gamma$ is similar to the dominating Higgs production process
$gg\to H$, both involving gluon-gluon fusion into a quark loop
(Fig.~\ref{fig-diagrams}). The experimental observation of prompt photon pairs
in rapidity gap events can therefore be used to test the models for diffractive
Higgs production. We therefore give predictions on such prompt photon events
based on our model. 

\section{Diffractive hard scattering in the SCI model}

The soft color interaction (SCI) model \cite{SCI} was developed in an attempt to
better understand non-perturbative QCD dynamics and provide a unified
description of all final states. The basic assumption is that soft color
exchanges give variations in the topology of the confining color string-fields
which then hadronize into different final states, \eg \ with and without
rapidity gaps or leading protons. Also other kinds of experimental results are
described in a very economical way with only one new parameter. Particularly
noteworthy is the turning of a $c\bar{c}$ pair from a color octet state into a
singlet state producing charmonium \cite{charmonium} in good agreement with
observed rates.

\begin{figure}[t]
\begin{center}
\epsfig{width= 0.9\columnwidth,file=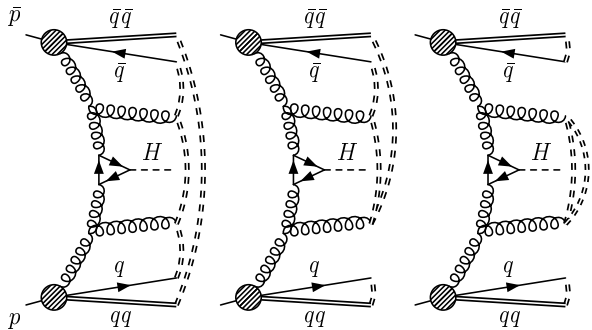,clip=}
\caption{Higgs production in $p\bar{p}$ collisions with string topologies
(double-dashed lines) before and after soft color interactions in the SCI model,
resulting in events with one or two rapidity gaps (leading protons).}
\label{pp-higgs}
\end{center}
\end{figure}

The SCI model \cite{SCI} is implemented in the Lund Monte Carlo programs \Lepto
\ \cite{Lepto} for deep inelastic scattering and \Pythia \ \cite{Pythia} for
hadron-hadron collisions. The hard parton level interactions are given by
standard perturbative matrix elements and parton showers, which are not altered
by the softer non-perturbative effects. The SCI model then applies an explicit
mechanism where color-anticolor (corresponding to non-perturbative gluons) can
be exchanged between the emerging partons and hadron remnants. The probability
for such an exchange cannot be calculated and is therefore taken to be a
constant given by a phenomenological parameter $P$. These color exchanges
modify the color connections between the partons and thereby the color
string-field topology, as illustrated in Fig.~\ref{pp-higgs}. Standard Lund
model hadronization \cite{lund} of the string fields then leads to different
final states, with gaps in rapidity regions where no string was present.
Following the normal factorization theorem, these soft processes do not affect
the cross section for the hard scattering process, but only the distribution of
hadrons in the final state. 

A variation is provided by the generalized area law (GAL) model \cite{GAL},
which is formulated in terms of interactions between the strings and not the
partons. The probability for two strings to interact via a soft color exchange
is obtained from the area law in the Lund string model resulting in a 
dynamically varying interaction probability. The results of this model are in
practice very close to those of the SCI model and we will therefore in this
short note only consider the SCI model. 

Since the SCI model is implemented in Monte Carlo programs which generate
complete events with final state particles, one can adopt an experimental
approach to classify events depending on the final state, \ie \ diffractive or
non-diffractive depending on the presence of rapidity gaps or leading
(anti)protons. The value of the single parameter $P$, regulating the amount of
soft color exchanges, is determined from the rate of diffractive deep inelastic
events observed in $ep$ at HERA. The predictive power of the model lies in the
fact that applying it, with the same parameter value ($P=0.5$), to \ppbar \
gives a good description of {\em all} diffractive hard scattering processes
observed at the Tevatron, as summarized in Table~\ref{tab-gapratios}. 

The \Pythia \ Monte Carlo has here been employed for different hard subprocesses
producing jets, $W$, $Z$, $c\bar{c}$ or $b\bar{b}$. The overall cross sections
are obtained as usual by folding the subprocess cross sections with the parton
density distributions (we have used CTEQ \cite{CTEQ} as detailed in
\cite{SCI-TEV}). The SCI model is then applied before hadronization is
performed and the events classified. Single diffractive (SD) events are
selected using one of two criteria: (1) a leading (anti)proton with $x_F>0.9$
or (2) a rapidity gap in the forward or backward region, for example
$2.4<|\eta|<5.9$ as used by the CDF collaboration. Applying the conditions in
both hemispheres results in events with two rapidity gaps, one on each side of
the central hard scattering system. These are conventionally labeled DPE after
their interpretation as double pomeron exchange in the Regge framework, but we
prefer to use this established acronym to denote `Double leading Proton Events'
independently of interpretations in terms of any specific model. 

\begin{table}
\caption{Ratios diffractive/inclusive for hard scattering processes in \ppbar \
collisions ($\sqrt{s}=1800$ GeV) at the Tevatron, showing experimental results
from CDF \cite{CDF,CDF-DPE} and D\O{} \cite{D0} compared to the soft color
interaction (SCI) model.}
\label{tab-gapratios}
\begin{tabular}{lllc}
\hline
\hline
             & \multicolumn{3}{c}{Ratio [\%]} \\
Observable \hspace*{10mm}   & \multicolumn{2}{c}{Observed} & SCI model \\
\hline
$W$  - gap       & CDF & $1.15 \pm 0.55$       & 1.2\\
$Z$  - gap       & D\O & $1.44^{+0.62}_{-0.54}$& 1.0\footnotemark[1] \\
$b\bar b$  - gap & CDF & $0.62 \pm 0.25$       & 0.7 \\
$J/\psi$  - gap  & CDF & $1.45 \pm 0.25$       & 1.4\footnotemark[1] \\
$jj$ - gap       & CDF & $0.75 \pm 0.10$       & 0.7 \\
$jj$ - gap       & D\O & $0.65 \pm 0.04$       & 0.7 \\
gap - $jj$ -gap\footnotemark[2] & CDF & $0.26 \pm 0.06$   & 0.2 \\
$\bar{p}$ - $jj$ -gap\footnotemark[2] & CDF & $0.80 \pm 0.26$   & 0.5 \\ 
\hline
\multicolumn{4}{l}{\footnotemark[1]{~Prediction of model}}\\
\multicolumn{4}{l}{\footnotemark[2]{~Ratio DPE/SD}}\\
\hline
\hline
\end{tabular}
\end{table}

For a detailed comparison of the SCI model with results on diffractive hard
scattering at the Tevatron see \cite{SCI-TEV}. Here we want to point out that
some of our results were predictions that have recently been verified by new
experimental data, as indicated in Table~\ref{tab-gapratios}. 

A process which has drawn much attention recently is the central production
of dijets in DPE events, which has been observed at the Tevatron~\cite{CDF-DPE}.
The SCI model describes well the cross section and the ratio of these
events to single diffractive dijet events (as shown in Table~\ref{tab-gapratios}),
which indicates the breakdown of diffractive factorization~\cite{CDF-DPE,SCI-TEV}.
Also more exclusive quantities, such as transverse energy of the jets or
their rapidity distribution are well described~\cite{SCI-TEV}.

\begin{figure}[bhtp]
\begin{center}
\epsfig{width= 0.95\columnwidth,file=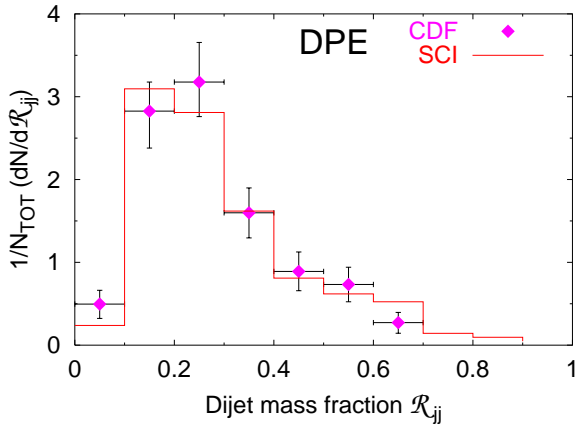}
\caption{Distribution of the dijet mass fraction, \ie \ ratio ${\cal R}_{jj}$ of
the invariant mass of the dijet system to the invariant mass of the central
hadronic system in DPE events. CDF data \protect\cite{CDF-DPE} compared to the
SCI model.}
\label{massfraction}
\end{center}
\end{figure}

In Fig.~\ref{massfraction} we show that another new measurement is well
described by the SCI model. This dijet mass fraction ${\cal R}_{jj}$ is an
interesting observable that shows how large fraction of the available energy in
the central system of DPE events that goes into the hard scattering system and
produces the two jets in the final state. Events with ${\cal R}_{jj}$ close to
unity would correspond to an exclusive jet production process and the data
limits such a contribution to $\sigma \lesssim 3.7$ nb \cite{CDF-DPE} just
above a model prediction of $\sim 1$ nb \cite{KMR}. The observed dijet mass
fraction is instead concentrated at smaller values, leaving a substantial
energy fraction in the remaining central DPE system, in accordance with the jet
production mechanism in the SCI model. 

Turning now to Higgs production, we select those hard subprocesses in \Pythia \
which  produce a Higgs boson. The dominant one is $gg\to H$ via a quark loop as
illustrated in Fig.~\ref{fig-diagrams}a, which accounts for 50\% and 70\% of
the cross section (for $115<m_H<200$ GeV) at the Tevatron and LHC,
respectively. Other production channels ($q_i\bar{q}_i\to H$, 
$q_i\bar{q}_i\to Z\, H$, $q_i\bar{q}_j\to W\, H$, $q_iq_j\to q_kq_l H$ and 
$gg\to q_k \bar{q}_k\, H$) contribute, depending both on the Higgs mass and the
center of mass energy. Our results in the following are based on these leading
order processes, but we note that higher order corrections could give an
increase of the Higgs production cross section with an effective $K$-factor of
order two~\cite{K-factor}.

Applying the same SCI model on the resulting partonic state gives rise to
different color string topologies, resulting in different final states after
the standard Lund model \cite{lund} has been applied for hadronization
(Fig.~\ref{pp-higgs}). Classifying events as above, depending on rapidity gaps
or leading protons, we obtained the results for diffractive Higgs production as
shown in detail in \cite{PRL-Higgs}. We summarize the main results in terms of
overall cross sections in Table~\ref{tab-higgs}. 

At the Tevatron, the cross section for Higgs in DPE events, which would have the
least disturbing underlying hadronic activity, is too small to give any
produced events. Higgs in single diffractive events are produced, but in small
numbers such that only the decay mode $H\to b\bar{b}$ with the largest
branching ratio will give any events to search for.
\begin{table}[t]
\caption{Cross sections and numbers (\#) of events at the Tevatron and LHC for
Higgs in single diffractive (SD) and DPE events, defined by leading protons or
rapidity gaps, obtained from the soft color interaction model (SCI).
\label{tab-higgs}}
\begin{tabular}{lllcc}
\hline
\hline
        & & & Tevatron & LHC \\
\multicolumn{3}{l}{$m_H=115$~GeV} & $\sqrt{s}=1.96$~TeV & $\sqrt{s}=14$~TeV \\
        & & & ${\cal L}=20~\mbox{fb}^{-1}$ & ${\cal L}=30~\mbox{fb}^{-1}$ \\
\hline
&$\sigma [\mbox{fb}]$ Higgs-total   & & {$600$}        & {$27000$} \\
\hline 
SD &$\sigma \; [\mbox{fb}]$ leading-p  & & $1.2$  & $190$  \\
   &$\sigma \; [\mbox{fb}]$ gap       &  & $2.4$ & $27$    \\
   &\# H + leading-p                   & & $24$ & $5700$  \\
   &$\hookrightarrow$ \# H $\rightarrow \gamma\gamma$  && $0.05$  & $13$ \\
\hline 
DPE &$\sigma \; [\mbox{fb}]$ leading-p's& &  $1.2 \cdot 10^{-{4}}$  &  $0.19$  \\
    &$\sigma \; [\mbox{fb}]$ gaps      & & $2.4 \cdot 10^{-{3}}$  &  $2.7\cdot 10^{-{4}}$   \\
    &\# H + leading-p's                             & & $0.0024$          & $6$  \\        
\hline
\hline
\end{tabular}
\end{table}

At LHC, the high energy and luminosity facilitates a study of single diffractive
Higgs production, where also the striking $H\to \gamma \gamma$ decay should
be observed. Also a few DPE Higgs events may be observed, but these events will not be as clean as naively expected. The available energy is here enough to produce the Higgs and the leading protons as well as an underlying event that will populate forward detector rapidity regions with particles \cite{PRL-Higgs}.
This causes a much lower diffractive cross section when requiring a gap instead
of a leading proton at LHC. 


\section{$H\to\gamma\gamma$ versus prompt $\gamma\gamma$}

The $H\to\gamma\gamma$ decay mode is of experimental interest due to its clear
signature with two photons of high $p_\perp$ (up to $m_H/2$). Its branching
ratio is, however, quite low since it proceeds via a higher order loop diagram
(Fig.~\ref{fig-diagrams}b) and is only of interest for an intermediate mass
Higgs, below the thresholds for decaying into $W^+W^-$ and $Z^0Z^0$. As the
Higgs mass increases from $m_H =115$ GeV to $m_H=160$ GeV, the branching ratio
decreases from about  
$2\times 10^{-3}$ to about $6\times 10^{-4}$ \cite{Higgs-BR}. Therefore, this
decay mode gives too low rates to be observable in diffractive interactions
(Table~\ref{tab-higgs}), except for a handful $H\to\gamma\gamma$ in single
diffraction at LHC. 

\begin{figure*}[htbp]
\begin{center}
\epsfig{width= 1\columnwidth,file=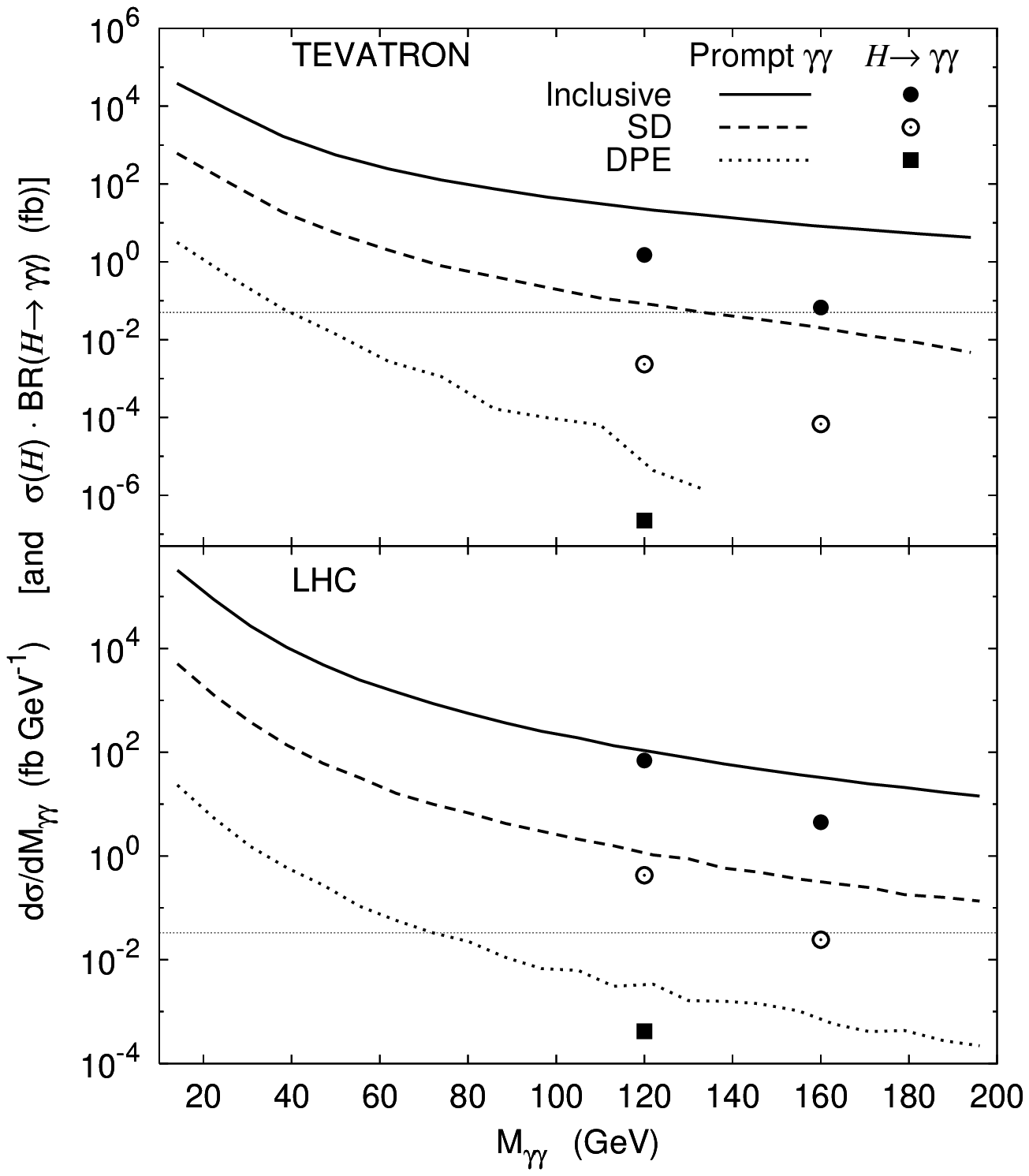}
\epsfig{width= 1\columnwidth,file=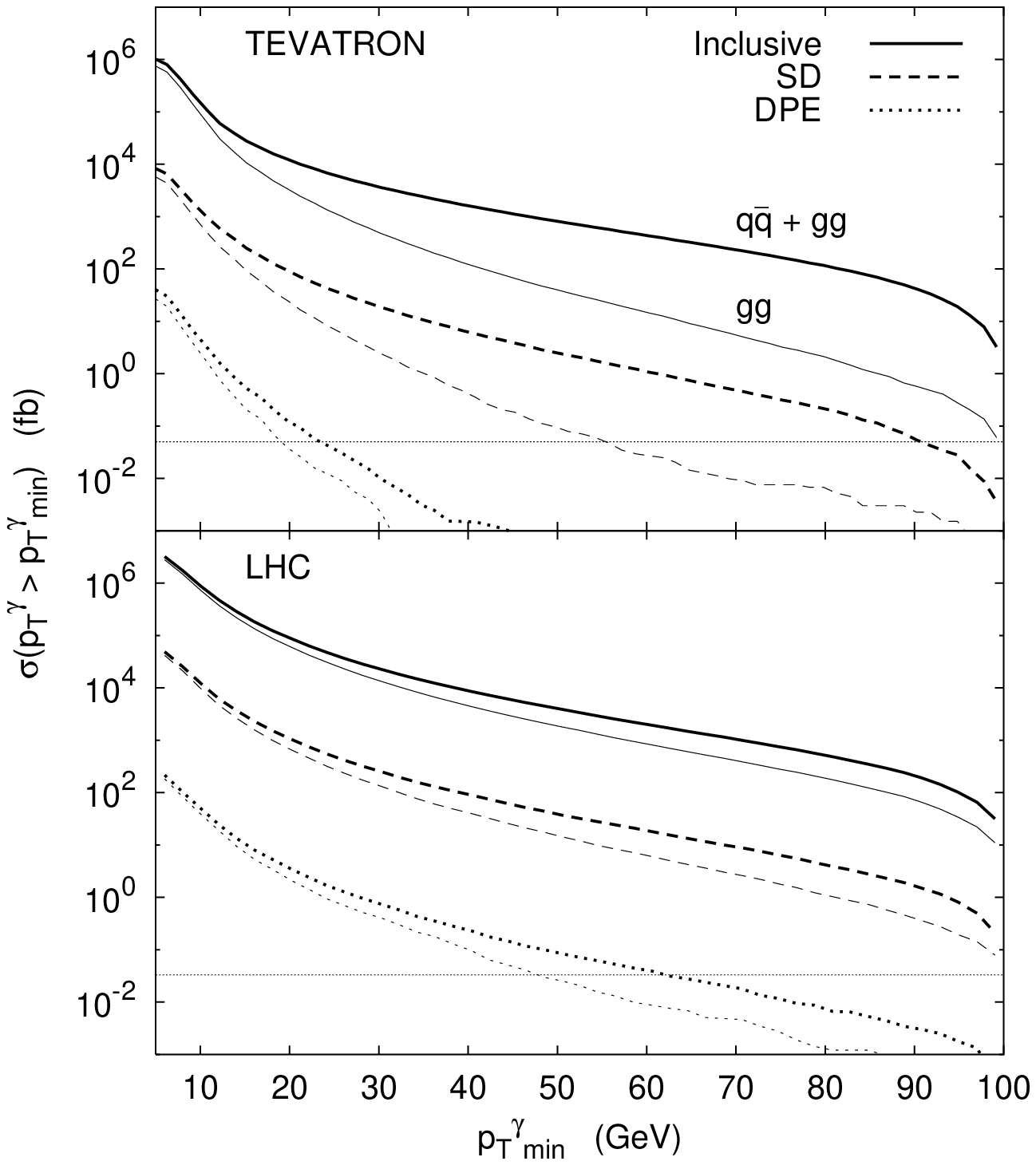}
\caption{Cross sections at the Tevatron and LHC for the production of prompt
$\gamma\gamma$ (with $|\eta_\gamma |<2$) versus (a) their invariant mass
$M_{\gamma\gamma}$ and (b) integrated from a minimum transverse momentum
$p^\gamma_{\perp {min}}$. Curves are predictions from the soft color
interaction model; inclusive, single diffractive and DPE events and in (b)
showing separately the contribution from the $gg\to\gamma\gamma$ process (thin
lines). For comparison, $\sigma(H)\cdot\rm{BR}(H\to\gamma\gamma)$ is shown in
(a) for $m_H=120$~GeV ($\mathrm{BR}=2.2\times 10^{-3}$) and $m_H=160$~GeV
($\mathrm{BR}=5.5\times 10^{-4}$). The horizontal lines show the cross-section for
obtaining
one event with the planned integrated luminosities of 20 and 30~fb$^{-1}$ at the
Tevatron and LHC, respectively.}
\label{fig-gammagamma}
\end{center}
\end{figure*}

Concerning the backgrounds to this Higgs signal, the requirement of the two
high-$p_\perp$ photons to be essentially back-to-back in the transverse plane
and isolated, removes the major backgrounds of photons as decay products in
jets. The serious, irreducible background is given by the production of a pair
of prompt high-$p_\perp$ photons from the hard processes
$q\bar{q}\to\gamma\gamma$ and $gg\to\gamma\gamma$
(Fig.~\ref{fig-diagrams}cd). As above, we simulate these processes with \Pythia
\ including the SCI model and obtain the inclusive cross sections, as well as
the single diffractive and DPE ones shown in Fig.~\ref{fig-gammagamma}. 
The Tevatron cross sections for diffractive Higgs and prompt photons with high
mass are very suppressed due to kinematics at the edge of phase space, while at
the LHC, this kinematic suppression is much less pronounced.

As can be seen, the irreducible prompt $\gamma\gamma$ background is larger,
in particular at the Tevatron, than the cross section times the branching ratio
for
$H\to\gamma\gamma$. This applies to inclusive, SD and DPE events when
considering a bin width of only $\sim$1~GeV, corresponding to the width of
a Higgs with $m_H=160$~GeV and a detector with good energy resolution.
For a Higgs of lower mass the width is much smaller, \eg{} $\sim$10~MeV at
$m_H=120$~GeV, such that the $H\to\gamma\gamma$ signal could in priciple
stand out from the background in 10~MeV bins obtained from a detector with
extremely high energy resolution. To be realistic, however, we must unfortunately
conclude that $\gamma\gamma$ is not a straightforward signal for observing
the Higgs boson.

On the other hand, the prompt $\gamma\gamma$ production has a large enough cross
section to be investigated in diffractive events. 
At the Tevatron, observable rates are predicted for SD events with
$p_\perp^\gamma$ up to $\sim 75$ GeV, so the model can be tested against data
even up to scales of order $m_H/2$, but for DPE, $p_\perp^\gamma$ is only
observable up to 15-20 GeV.
For large $p_\perp^\gamma$ the leading order subprocess
$q\bar{q}\to\gamma\gamma$ dominates. The higher order process
$gg\to\gamma\gamma$ dominates, however, at lower $p_\perp^\gamma$ 
where the momentum fraction $x$ of the initial partons can be smaller and the
large gluon density of the proton increases the cross section. Therefore, in
order to test the model for a similar gluon-gluon fusion process with a quark
loop as in Higgs production, one should consider $p_\perp^\gamma$ up to $\sim
15$ GeV for SD  and up to $\sim 20$ GeV for DPE (Fig.~\ref{fig-gammagamma}b).

Similarly at the LHC, prompt $\gamma\gamma$ gives observable rates for SD for
$p_\perp^\gamma$ up to $\sim 100$ GeV and for DPE up to $\sim 50$ GeV, \ie \ up
to the scale of $m_H/2$. The process $gg\to\gamma\gamma$ dominates for values
of $p_\perp^\gamma$ below $\sim 60$ GeV for SD and below $\sim 40$ GeV for DPE.

Thus, the basic mechanism for producing rapidity gap events with a high mass
state from $gg$ fusion via a quark loop can be tested already at the Tevatron
and further investigated at the LHC. 


\section{Concluding discussion}

Diffractive Higgs production is currently considered with great interest for
forthcoming investigations at the Tevatron and LHC. Different theoretical
models predict quite different magnitudes of the cross section, as discussed
recently \cite{Khoze:2002py,DeRoeck:2002pr}. In this context, diffractive
prompt $\gamma\gamma$ production can be used as a further testing ground for
the models. In particular, the contributing subprocess $gg\to\gamma\gamma$
involves a similar gluon-gluon fusion process via a quark loop as in Higgs
production.

Although models based on pomeron exchange can be made to fit diffractive hard
scattering data at both HERA and the Tevatron, they have conceptual and
theoretical problems as discussed in \cite{SCI-TEV,Brodsky-Hoyer}. In
particular, it seems improper to regard the pomeron as `emitted' from the
proton and having QCD evolution as a separate entity. This problem is avoided
in the soft color interaction model which, although not having a firm
theoretical basis, can be seen as an effective model for final state
interactions as discussed theoretically in \cite{Brodsky-Hoyer}. This SCI model
has, furthermore, strong phenomenological support since it has been very
successful in describing and predicting diffractive hard scattering processes
both in $ep$ collisions at HERA and \ppbar \ at the Tevatron \cite{SCI-TEV}. 
It should be emphasized that the SCI model describes well the production of
dijets in DPE events -- a process which has similar dynamics to DPE Higgs and
has been advocated~\cite{DeRoeck:2002pr} as a testing ground for different
models aiming at describing diffractive Higgs.

All this gives a high credibility to our predictions for diffractive Higgs
\cite{PRL-Higgs} and prompt $\gamma\gamma$ presented in this note. 

Based on the SCI model, we find that the rate of diffractive Higgs events at the
Tevatron will be too low to be useful. Some single diffractive Higgs events
should be produced, but they can only be accessed through the $H\to b\bar{b}$
decay channel which does not provide a clear signal in view of the large QCD
background and non-trivial experimental $b\bar{b}$ reconstruction. The Higgs
must, therefore, be primarily searched for in inclusive events with their more
complex underlying event. The situation is quite different at LHC. Here single
diffractive Higgs production can be investigated, including a few events with
the striking $H\to\gamma\gamma$ decay mode, and a few DPE events with a Higgs
should also be produced. Diffractive events are, however, not so clean at the
LHC since the large available energy produces underlying event activity
extending to larger forward rapidities making observable gaps less abundant.
Therefore, a much larger sample of diffractive Higgs events would be obtained
by very forward `Roman pot'-type proton tagging detectors. 

We furthermore find that the $H\to\gamma\gamma$ signal in all cases has a
smaller cross section than the irreducible background of prompt $\gamma\gamma$
production. As discussed above, the latter process is interesting in its own
right and our model gives observably large cross sections both at the Tevatron
and LHC, such that prompt $\gamma\gamma$ event samples should be obtained
inclusively as well as in single diffraction and DPE. A one to two orders of
magnitude larger cross section for prompt $\gamma\gamma$ in DPE was predicted
in \cite{Cox} based on a pomeron model, but the uncertainty was considered
large due to the poorly constrained Reggeon exchanges. Thus, the investigation
of diffractive prompt $\gamma\gamma$ production at the Tevatron is essential to
test the models for diffractive hard scattering and obtain safer predictions
for diffractive Higgs production at the LHC. 



\end{document}